\documentclass[12pt]{article}\usepackage[]{graphicx}\usepackage[]{color}
\makeatletter
\def\maxwidth{ %
  \ifdim\Gin@nat@width>\linewidth
    \linewidth
  \else
    \Gin@nat@width
  \fi
}
\makeatother

\definecolor{fgcolor}{rgb}{0.345, 0.345, 0.345}
\newcommand{\hlnum}[1]{\textcolor[rgb]{0.686,0.059,0.569}{#1}}%
\newcommand{\hlstr}[1]{\textcolor[rgb]{0.192,0.494,0.8}{#1}}%
\newcommand{\hlopt}[1]{\textcolor[rgb]{0,0,0}{#1}}%
\newcommand{\hlstd}[1]{\textcolor[rgb]{0.345,0.345,0.345}{#1}}%
\newcommand{\hlkwb}[1]{\textcolor[rgb]{0.69,0.353,0.396}{#1}}%
\newcommand{\hlkwc}[1]{\textcolor[rgb]{0.333,0.667,0.333}{#1}}%
\newcommand{\hlkwd}[1]{\textcolor[rgb]{0.737,0.353,0.396}{\textbf{#1}}}%

\usepackage{framed}
\makeatletter
\newenvironment{kframe}{%
 \def\at@end@of@kframe{}%
 \ifinner\ifhmode%
  \def\at@end@of@kframe{\end{minipage}}%
  \begin{minipage}{\columnwidth}%
 \fi\fi%
 \def\FrameCommand##1{\hskip\@totalleftmargin \hskip-\fboxsep
 \colorbox{shadecolor}{##1}\hskip-\fboxsep
     \hskip-\linewidth \hskip-\@totalleftmargin \hskip\columnwidth}%
 \MakeFramed {\advance\hsize-\width
   \@totalleftmargin\z@ \linewidth\hsize
   \@setminipage}}%
 {\par\unskip\endMakeFramed%
 \at@end@of@kframe}
\makeatother

\definecolor{shadecolor}{rgb}{.97, .97, .97}
\definecolor{messagecolor}{rgb}{0, 0, 0}
\definecolor{warningcolor}{rgb}{1, 0, 1}
\definecolor{errorcolor}{rgb}{1, 0, 0}
\newenvironment{knitrout}{}{} % an empty environment to be redefined in TeX

\usepackage{alltt}
\usepackage{amsmath}
\usepackage{graphicx,psfrag,epsf}
\usepackage{enumerate}
\usepackage[affil-it]{authblk}
\usepackage{graphicx}
\usepackage[space]{grffile}
\usepackage{latexsym}
\usepackage{textcomp}
\usepackage{longtable}
\usepackage{multirow,booktabs}
\usepackage{amsfonts,amsmath,amssymb}
\usepackage{url}
\usepackage{hyperref}
\hypersetup{colorlinks=false,pdfborder={0 0 0}}
\newif\iflatexml\latexmlfalse
\DeclareGraphicsExtensions{.pdf,.PDF,.png,.PNG,.jpg,.JPG,.jpeg,.JPEG}

\usepackage[utf8]{inputenc}
\usepackage[english]{babel}

\newcommand{\rpkg}{\texttt}
\newcommand{\rexpr}{\texttt}
\newcommand{\rfun}{\texttt}

\newcommand{\proglang}{\textbf}
\usepackage{natbib}
\bibpunct[, ]{(}{)}{;}{a}{}{,} %citation style for ASA style guide 

\newcommand{\blind}{0}

\IfFileExists{upquote.sty}{\usepackage{upquote}}{}
\begin{document}

\def\spacingset#1{\renewcommand{\baselinestretch}%
{#1}\small\normalsize} \spacingset{1}

%% \SweaveOpts{concordance=TRUE}

\if0\blind
{

  \title{trackr: A Framework for Enhancing Discoverability and Reproducibility of Data 
    Visualizations and Other Artifacts in R}

  \author{Gabriel Becker \\
    Department of Bioinformatics and Computational Biology, Genentech Inc.\\
    and \\
    Sara E. Moore \\
    Department of Biostatistics, University of California, Berkeley \\
    and \\
    Michael Lawrence \\
    Department of Bioinformatics and Computational Biology, Genentech Inc.}
  \maketitle
} \fi

\if1\blind
{
  \bigskip
  \bigskip
  \bigskip
  \begin{center}
    {\LARGE\bf trackr: A Framework for Enhancing Discoverability and Reproducibility of Data 
    Visualizations and Other Artifacts in R}
\end{center}
  \medskip
} \fi

%%   \author[1]{Gabriel Becker}
%%   \author[2]{Sara A. Moore}
%%   \author[1]{Michael Lawrence}
  
%% \affil[1]{Dept of Bioinformatics and Computational Biology, Genentech}
%% \affil[2]{Dept of Biostatistics, University of California, Berkeley}

%% \date{\today}

%% \maketitle

\begin{abstract}
  Research is an incremental, iterative process, with new results
  relying and building upon previous ones. Scientists need to find,
  retrieve, understand, and verify results in order to confidently
  extend them, even when the results are their own. We present the
  \rpkg{trackr} framework for organizing, automatically annotating,
  discovering, and retrieving results. We identify sources of
  automatically extractable metadata for computational results, and we
  define an extensible system for organizing, annotating, and
  searching for results based on these and other metadata. We present
  an open-source implementation of these concepts for plots,
  computational artifacts, and woven dynamic reports generated in the
  R statistical computing language.
\end{abstract}

\noindent%

\newpage
\spacingset{1.45} % DON'T change the spacing!

\section{Introduction}

Research is an incremental, iterative process, with new results
relying and building upon previous ones. Scientists need to find,
retrieve, understand, and verify results in order to confidently
extend them, even when the results are their own. Understanding an
existing result requires knowing how it was generated
\citep{rossini2003literate,becker2014dynamic}. Associating a result
with the set of actions \citep{Lee_1995} or code
\citep{Claerbout_1992,Buckheit_1995,Leisch_2002} that generated it
enhances the understandability and reproducibility of the result.
Further associating a result with information about the input data and
the software involved provides even stronger guarantees of
reproducibility \citep{Gentleman_2004,Stodden_2015}.  Researchers must
be aware of a result in order to extend it, so the tracking,
organization, and discoverability of results is key to their
incorporation into a larger body of scientific work.  The ability of a
scientist to find and retrieve an existing result from within a body
of results depends on how well the result is annotated with metadata
that enable its discovery and differentiate it from other results
\citep{figshare_faq}.

Standardized tracking, organization, and annotation of results serves different 
purposes when leveraged at different organizational scales. Individual analysts 
often generate dozens of visualizations and other artifacts during a single 
analysis. They then retrieve particular results as needed --- e.g., when 
presenting in a group meeting, preparing a manuscript, or reviewing past work. 
When their results are organized and stored in standard ways, locating and 
retrieving the correct artifact is easier and less error-prone
\citep{Schwab_2000}. Searching for a result based on metadata annotations, for 
example the datasets and variables used, plot title, statistical model type and 
specification, etc., is generally easier still and more efficient than manually 
locating an image file or serialized result on the file system --- particularly 
as the number of results and the time since their creation grow.  Beyond 
locating individual results, metadata can identify groups of 
\emph{related} results, such as those involving a certain variable or data 
subset. This grants the analyst a more holistic, retrospective view of their 
analysis, potentially leading to new insights derived from the synergy of many 
individual results. 

Systematic tracking of results provides additional benefits to
research teams and the broader scientific community at large. Tracking
artifacts in a standardized way across multiple analysts, projects,
and research efforts facilitates the creation and effective use of a
body of institutional or communal knowledge. This increases an
organization's effectiveness by reducing duplication of work, helping
identify connections between related or synergistic work, and
increasing the impact of results. Sharing results is necessary for
collaboration between analysts and subject-material experts. For these
and other reasons, standalone publication of datasets, plots
\citep{plotly,shiny_showcase} and other computational artifacts
\citep{figshare,rpubs} is coming to be an accepted means of scientific
discourse.

Tracking results requires bookkeeping that distracts scientists from
pursuing answers to their scientific questions, so scientists benefit
from software that automatically annotates results with metadata
useful for rediscovery and stores the results in a queryable
database. \citet{archivist}'s \rpkg{archivist} \proglang{R} package
seeks to address some aspects of the problem. It provides a framework
for caching and (re-)discovering \proglang{R} objects based on manual
annotations and a limited number of automatic ones.

Recognizing the need for more comprehensive automatic annotations, we
present the \rpkg{trackr} and \rpkg{histry} \proglang{R}
packages. Together, these define a framework for for tracking,
automatically annotating, discovering, and reproducing the
intermediate and final results of computational work done within the
\proglang{R} statistical programming language \citep{Rlang}. We have
developed several interfaces to \rpkg{trackr}, for different use
cases, all of which integrate with existing R-based analysis
workflows. More generally, we have determined useful types of metadata
and how we can automatically generate them ---
particularly in the case of data visualizations.

\section{Methods}

Our \rpkg{trackr} framework aims to provide a flexible mechanism for
organizing, locating, and retrieving comptuational results and
information about them. To do this we will provide two key features:
the ability to \emph{record} a result so that it will be
annotated and stored in a way that ensures users will be able to
locate it, and a mechanism for automatically generating useful
metadata annotations to be associated with results.

Recording a result using the \rpkg{trackr} framework is an extension
of saving it to disk. When given a result, the \rpkg{trackr}
recording process:
\begin{enumerate}
  \item{saves the result in whatever form(s) \rpkg{trackr} is
    configured to retain,}
  \item{automatically infers metadata about it, and }
  \item{adds a record representing the result and its metadata to a
    searchable database of results \rpkg{trackr} knows about.}
\end{enumerate}
Users are then able to locate \emph{results} --- of their own and/or
of others, depending on the database
--- by searching their \emph{metadata}.

While our system is general and supports annotating and recording
arbitrary in-session computational artifacts, we have particularly
focused on extracting rich metadata from data visualizations. For the
remainder of this section, we discuss \emph{types} and
\emph{particular pieces} of metadata that we have identified as both
useful and automatically derivable from available information without
requiring manual input from the analyst. 

This section discusses metadata and its capture at a conceptual
level. We provide a concrete example of exactly what \rpkg{trackr}
captures in Section \ref{trackplot} and along with a complete metadata
record in the supplimentary materials.

\subsection{Identifying and computing useful artifact metadata} \label{art_metadata}

Certain pieces of low- to medium-level information provide insight into the 
purpose of a computation and the conclusions drawn from the result. These 
details are useful axes of inquiry when searching for the artifact or similar or
related ones from among a host (database) of others. 

The \rpkg{trackr} package constructs metadata annotations falling into four 
broad categories:
\begin{enumerate}
    \item \label{mdatsrc:1} Information about the computational environment 
      hosting \rpkg{trackr}
      \footnote{this is likely, though not guaranteed, to be the same 
        environment used to create 
        the artifact},
    \item \label{mdatsrc:2} Aspects of the artifact's design and/or structure,
    \item \label{mdatsrc:3} Aspects of the code that generated the artifact, 
      including upstream  analysis (when available), and 
    \item \label{mdatsrc:4} Aspects of the data (when available).
\end{enumerate}

\subsubsection{Information about the computing environment} \label{meth_comp_env}

When analysts use \rpkg{trackr} to record and track an artifact, we collect contextual 
information, including the submitting user and the timestamp of the submission. 
When storing the artifact in the database, we assign it a unique\footnote{up 
  to the collision probability of Jenkin's SpookyHash algorithm, roughly 
  $\frac{1}{2^{128}}$ \citep{jenkins2012spookyhash}} identifier using 
\citet{fastdigest}'s \rpkg{fastdigest} package. In addition, we capture 
information about the containing project and/or file when possible --- 
currently, when the code is being evaluated from a file within the RStudio 
IDE, or when \proglang{R} is running from within an \proglang{R} package 
directory structure.

\subsubsection{Aspects of the artifacts's design and construction} \label{meth_plot_des_constr}

A computational artifact is intrinsically linked to the question(s) the analyst
seeks to answer.  The analyst designs a statistical model or plot that aims to
answer the question, then writes and runs code to create the corresponding 
artifact. By capturing information about the underlying design of the artifact, 
we enable users to infer the questions motivating the analysis, locate
particular artifacts or broader classes thereof, and better interpret the 
results these artifacts represent. For example, an analyst wishing to compare
subpopulation distributions might generate a boxplot, based on the decision that
a boxplot would effectively compare the distributions. If we learn that the plot
is a boxplot, we help the discoverer infer that the analyst was comparing 
distributions. This is particularly effective when combined with other metadata
discussed in subsequent sections, such as the name and attributes of the dataset
being analyzed.

\citet{Wilkinson_1999} and \citet{ggplot} defined a grammar of graphics for
describing statistical plots. While intended and used primarily to \emph{create}
plots, we instead \emph{analyze} the formal description to make inferences about
a plot's content and purpose. For example, the set of variables being plotted,
represented by the aesthetic mappings, is a clue about the nature of the data
and the questions asked of it. Components like the geometric representation 
(\textit{geom}) and statistical transformation (\textit{stat}) are also 
relevant. For example, a scatterplot generally looks at the relationship between
two continuous variables, while a quantile-quantile plot specifically compares
two distributions. \emph{Small multiple} plots \citep{tufte1983visual} --- 
generated via \emph{faceting} in \rpkg{ggplot2}'s grammar or \emph{conditioning}
in \rpkg{lattice} \citep{lattice} --- and plots with variables mapped to
\emph{aesthetics} such as color or plotting symbol indicate multivariate
relationships of interest. Guides, such as plot titles, axis labels and values,
and the textual components of legends, are semantically rich annotations that 
convey the context of the analysis in terms a human reader can understand. By
annotating a plot with the geoms, stats, guides and other components of its
grammatical description, \rpkg{trackr} is able to preserve information about
the plot known to the analyst at creation-time, information that often
indicates the intent of the analysis.

The plotting system used to create a plot --- e.g., \rpkg{ggplot2},
\rpkg{lattice}, or \rpkg{graphics} --- is also informative. When searching for a
particular plot, the user often has some notion as to the software used to
generate it. That might be due to memory of its overall appearance, or simply
because a user tends to prefer a particular package for a certain type of plot.

\rpkg{trackr} extracts metadata directly from \rpkg{ggplot2} and \rpkg{lattice}
plot objects. In the case of base graphics, \rpkg{trackr} mines the 
\rexpr{recordedPlot} object and analyzes the generating code. Those sources are
semantically poor relative to the high-level plot objects, so \rpkg{trackr} is
unable to extract as much information from base graphics. 

Many non-plot artifacts have analogous specifications which can inform
annotations. For example, we annotate the object representing the fit of a
generalized linear model with the model design (formula) and the link
function. From this information, we can derive which variables the analyst
thought were important, and which relationships between them were of interest. 
Furthermore, unlike plot objects, model fits contain labeled numerical results.
For example, we can annotate a model fit with the names of all the significant 
terms\footnote{We are consciously ignoring the obvious pitfalls of
  multiple-comparisons when searching for all models that found a particular
  variable to be significant.}.

\subsubsection{Aspects of analysis code} \label{aspects_of_code}

The code that creates an artifact also describes it in a low-level way.
Automatically inferring semantics from the code is challenging; however, it is
still worth saving the code, because it enables technical reproducibility, and
anyone who finds the code can make inferences about it. 

% ML: Does it work in the batch context?
%     I'm pretty sure the parser always drops comments.
% ML: I would probably drop "traditional".
Comments would be a very good source of information about the
analyst's intentions. Unfortunately in the \proglang{R} parser does
not retain comments passed to it. Thus when, e.g., sending lines from
a script file into R via an IDE, comments are not available as a
source of metadata.

The name of the function that constructs the object is often informative. For
example, the names 
of the \rfun{hist} and \rfun{histogram} functions in the \rpkg{graphics} and
\rpkg{lattice} packages, respectively, unambiguously indicate the type of plot
they produce. Similarly, the names of the, e.g.,  \rfun{lm}, \rfun{glm}, and
\rfun{prcomp} functions each indicate the nature and purpose of the objects that
they generate.  

The code leading up to artifact generation contains additional information. If
we see that the \rfun{plot} function is called on an object which was returned
by the \rfun{density} function, for example, we can infer that the plot is a 
density plot. Furthermore, we know that the analyst was investigating the
distribution of a particular variable, and that the variable was 
(approximately) continuous. We gain similar, perhaps even stronger, insight if
we can detect that, e.g, the residuals of a model created by previous code are
plotted in any of the common diagnostic manners. In this case, we might append
information about the model to that plot's metadata, leveraging information
about the larger analytic context. 

String constants are also highly informative;  file names, website urls, and key
function arguments are all often found as strings within the code. Discovering 
one or more particular artifacts generated during an analysis becomes much 
easier when armed with file names for the data or the knowledge that the
analyst filtered by a specific sample id.

Finally, for the sake of reproducibility, we associate an artifact with the the
full, runnable set of expressions that generated it. With this code and 
access to the correct data and software, users can recreate results and
potentially extend the archived analysis. 

We use \citet{codedepends}'s \rpkg{CodeDepends} package to perform static code
analysis and determine which expressions, variables, and string constants act as
inputs --- potentially via a chain of dependency --- to the result generation,
as well as the generating function and packages loaded by the code. This depends
on a mechanism for capturing the code, which we describe in Section 
\ref{capture_code}.

\subsubsection{Aspects of analyzed data}

Type and summary information about the data can provide insight into the type
and purpose of an artifact, and even hint at its conclusions in some cases.
Median information for each group in a boxplot, for example, can help a user 
locate a particular plot. Similarly, knowing the range of the plotted points can
identify scatterplots of variables on a domain of interest. Summaries of 
numerical and categorical variables can further differentiate between, e.g.,
plots which display different subsets of a dataset. 

The \rpkg{lattice} and \rpkg{ggplot2} plotting systems incorporate the plotted
data into the generated \rexpr{trellis} or \rexpr{ggplot} object, though in
different forms. Model fitting systems in \proglang{R}, including the \rfun{lm}
and \rfun{glm} functions, generally do the same. Locating the data in the case
of base graphics and general \proglang{R} object artifacts, on the other hand,
is more difficult; here we must rely on code analysis heuristics. 

We also enable the analyst to provide additional annotations. While we can 
automatically infer a great deal about an \proglang{R} object (and what it 
represents), its creator will understand the context, purpose, and ultimate
message of an artifact in a way that our automated systems can only approximate.

% Seems like this should come after the capture_code section.
\subsection{Tracking and annotating dynamic documents} \label{extract_dyndoc}

When communicating results, scientists often combine individual
artifacts, including plots, model fits, processed datasets, and the
like, in a \emph{woven report}. Woven reports are results themselves
and holistically represent an analysis by integrating renderings of
multiple individual sub-results, descriptive text providing context and
interpretation, and a subset of the code that generated the results.

One of the major sources of discoverability for a report is
association with its results, and vice versa. An end-user might
discover a report by locating a plot displayed within it and following
the association from that plot to the containing report. The opposite
workflow, where the user locates an artifact by first discovering the
containing report --- e.g., via a query based on the descriptive text
--- is also important.

\subsubsection{Sources of metadata for woven reports}

The information \rpkg{trackr} extracts for woven reports falls into
five broad categories:
\begin{enumerate}
  \item{the computational environment hosting \rpkg{trackr},}
  \item{the set of result artifacts the report conveys,}
  \item{the descriptive text of the report,}
  \item{the full body of code within the report, and}
  \item{any header material or other metadata encoded within the report.}
\end{enumerate}

We describe elsewhere how we capture information about the
computational environment, component artifacts and code. We discuss
the other aspects below.

\subsubsection{Information about the set of rendered results}

We associate metadata extracted from individual results to the
metadata for the parent report. We also summarize the \emph{set} of
results via information such as: how many results are displayed in the
report, how many of them are plots or other specific types, and even
whether they depend on each other.

\subsubsection{Information about the descriptive text}

The descriptive text within a woven report explicitly details 
the context, conclusions, interpretation, and even implications of
the analysis it embodies. Capturing these aspects, even imperfectly,
enhances discoverability, as we can
now ask for, e.g., the set of all reports involving a particular dataset
whose text indicates that a particular gene was upregulated or a particular
variable was found to be significantly associated with the response of 
interest. 

% What about inferring the strength of association between text and
% result by proximity within the document?
Currently, \rpkg{trackr} captures the text as a whole, allowing it to
be indexed and made searachable by whatever backend is in use. In
the future, we might extract additional semantics by inferring key
phrases and concepts within the text.

\subsubsection{Information from metadata within the dynamic document}

Dynamic document formats typically encode metadata including author,
title, destination format, and more \citep{rmarkdown}.  When
available, we annotate the report with that information. \proglang{R}
package vignettes also encode additional ``vignette metadata'',
including keywords, package dependencies required by the vignette,
etc.

%% Do we even need THIS section? does it really say anything? ~GB

\subsection{Tracking and incorporating the history of successful \proglang{R} expressions} \label{capture_code}
We showed in \ref{aspects_of_code} that the body of code used to generate an
object is a rich source of information about its structure and purpose. For
convenience and fidelity, we automatically capture the code behind an artifact,
whether generated in batch mode or interactively. Our algorithm for 
automatically associating code with an artifact has two steps:

\begin{enumerate}
    \item \label{track_code1} Capture and retain all \emph{successfully 
      evaluated} top-level expressions within the interactive session or 
      document being processed, and
    \item \label{track_code2} identify the subset of expressions relevant to the
      creation of a specified artifact.
\end{enumerate}

For (1), we provide a formal abstraction for keeping a running history of
successfully evaluated expressions in the \rpkg{histry} package. We have implemented the abstraction for
interactive use and for \rpkg{knitr} reports.

For (2), we again use Temple Lang et al.'s \rpkg{CodeDepends} to
identify expressions that contributed to the creation of the
artifact. This is the same set of expressions that would be required
to regenerate the artifact, to the extent that the code was
recorded. In the ideal case this code encompasses the entire program
for creating the artifact, starting from the data, assuming they have
not changed.  We then extract the metadata we described in Section
\ref{aspects_of_code} from the filtered expressions.

\section{Results}

In this section, we describe \rpkg{trackr} as software, including its user 
interface and how it can be extended. Full code for the plots and R operations
in these sections are included in the supplemental materials.

\subsection{Annotating and tracking an \proglang{R} plot with \rpkg{trackr}} \label{trackplot}

Tracking a plot with \rpkg{trackr} requires two inputs:
\begin{enumerate}
\item An \proglang{R} object representing the artifact; and,
\item A description of where/how \rpkg{trackr} should store artifacts and their
  metadata.
\end{enumerate}

Suppose we start with a plot of 3000 observations sampled from the
\rexpr{diamonds} dataset provided by \citet*{ggplot}'s \rpkg{ggplot2} package:

\begin{knitrout}
\definecolor{shadecolor}{rgb}{0.969, 0.969, 0.969}\color{fgcolor}\begin{kframe}
\begin{alltt}
\hlkwd{library}\hlstd{(trackr)}
\hlkwd{library}\hlstd{(ggplot2)}
\hlkwd{set.seed}\hlstd{(}\hlnum{620}\hlstd{)}
\hlstd{dsamp} \hlkwb{<-} \hlstd{diamonds[}\hlkwd{sample}\hlstd{(}\hlkwd{nrow}\hlstd{(diamonds),} \hlnum{3000}\hlstd{), ]}
\hlstd{d} \hlkwb{<-} \hlkwd{ggplot}\hlstd{(dsamp,} \hlkwd{aes}\hlstd{(carat, price,} \hlkwc{colour} \hlstd{= clarity))} \hlopt{+}
    \hlkwd{geom_point}\hlstd{()} \hlopt{+} \hlkwd{geom_smooth}\hlstd{(}\hlkwc{se} \hlstd{=} \hlnum{FALSE}\hlstd{)} \hlopt{+}
    \hlkwd{scale_colour_manual}\hlstd{(}\hlkwc{values} \hlstd{=} \hlkwd{c}\hlstd{(}\hlstr{"#999999"}\hlstd{,} \hlstr{"#E69F00"}\hlstd{,} \hlstr{"#56B4E9"}\hlstd{,}
                                   \hlstr{"#009E73"}\hlstd{,} \hlstr{"#F0E442"}\hlstd{,} \hlstr{"#0072B2"}\hlstd{,}
                                   \hlstr{"#D55E00"}\hlstd{,} \hlstr{"#CC79A7"}\hlstd{))}
\hlstd{d}
\end{alltt}

{\ttfamily\noindent\itshape\color{messagecolor}{\#\# `geom\_smooth()` using method = 'loess'}}\end{kframe}
\includegraphics[width=\maxwidth]{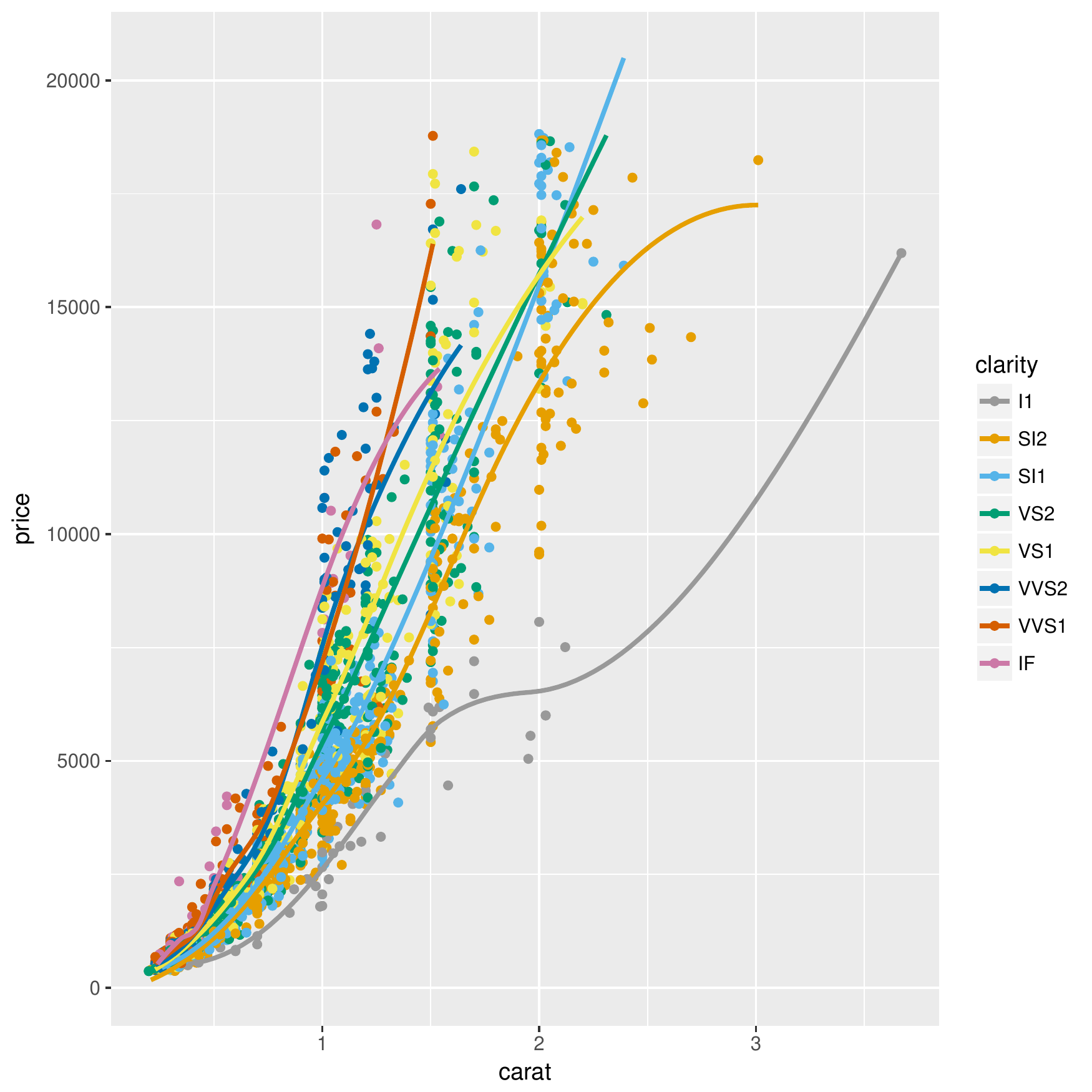} 

\end{knitrout}

By default, \rpkg{trackr} uses a backend based that writes to a local JSON file
in a standard, user-specific location. Our code below assumes that this backend 
is in use, or that the default backend has been replaced within the session. We 
refer readers to the package vignettes for details on how this would be done. 
We are now add our plot, \rexpr{d}, to the database by calling the \rfun{record}
function. 

\begin{knitrout}
\definecolor{shadecolor}{rgb}{0.969, 0.969, 0.969}\color{fgcolor}\begin{kframe}
\begin{alltt}
\hlkwd{record}\hlstd{(d)}
\end{alltt}

{\ttfamily\noindent\color{warningcolor}{\#\# Warning in if (as.character(fn) == "{}close"{}) \{: the condition has length > 1 and only the first element will be used}}

{\ttfamily\noindent\itshape\color{messagecolor}{\#\# `geom\_smooth()` using method = 'loess'\\\#\# `geom\_smooth()` using method = 'loess'\\\#\# `geom\_smooth()` using method = 'loess'}}\end{kframe}
\end{knitrout}

Once the artifact is in the database, we can search for it using 
\rfun{findRecords}, given a search pattern \footnote{often a regular 
  expression, though support for this technically depends on the backend in 
  use}.

\begin{knitrout}
\definecolor{shadecolor}{rgb}{0.969, 0.969, 0.969}\color{fgcolor}\begin{kframe}
\begin{alltt}
\hlstd{plt} \hlkwb{<-} \hlkwd{findRecords}\hlstd{(}\hlstr{"smooth"}\hlstd{,} \hlkwc{ret_type} \hlstd{=} \hlstr{"id"}\hlstd{)}
\hlstd{plt}
\end{alltt}
\begin{verbatim}
## [1] "SpkyV2_ce8a4f2e063ce6ca3c4a1e770cba1f8b"
\end{verbatim}
\end{kframe}
\end{knitrout}

We can see a subset of the specific metadata captured by rehydrating
and printing the \rexpr{FeatureSet} object corresponding to our plot,
like so:

\begin{knitrout}
\definecolor{shadecolor}{rgb}{0.969, 0.969, 0.969}\color{fgcolor}\begin{kframe}
\begin{alltt}
\hlstd{fs} \hlkwb{=} \hlstd{trackr}\hlopt{:::}\hlkwd{listRecToFeatureSet}\hlstd{(}\hlkwd{findRecords}\hlstd{(plt,} \hlkwc{fields} \hlstd{=} \hlstr{"uniqueid"}\hlstd{)[[}\hlnum{1}\hlstd{]])}
\hlkwd{show}\hlstd{(fs)}
\end{alltt}
\begin{verbatim}
## An GGplotFeatureSet for a ggplot2 plot 
## id: SpkyV2_ce8a4f2e063ce6ca3c4a1e770cba1f8b 
## tags: ggplot gg, ggplot 
## location: 1 lines of code in <unknown file> within 
## 	rstudio project: NA 
## 	package: 
## titles: NA 
## vars: carat <x>, price <y>, clarity <group.color> 
## facets:
## geom(s): point smooth 
## stat(s): identity smooth
\end{verbatim}
\end{kframe}
\end{knitrout}

Here we have captured the \emph{geoms} and \emph{variables} used in
the construction of the plot, as well as some standard information
about where the plot resides.  More complete static code analysis
results and \rfun{sessionInfo()} output are also captured; these are
omitted by the \rfun{show} method for brevity. We refer interested
readers to the supplimentary materials for a complete JSON record
for this plot.

Finally, if we need to remove a plot from our database from within 
\proglang{R}, we can use the \rfun{rmRecord} function. This function 
can take the artifact being removed --- if available within the 
\proglang{R} session --- or the unique ID associated with it in the 
database.

\begin{knitrout}
\definecolor{shadecolor}{rgb}{0.969, 0.969, 0.969}\color{fgcolor}\begin{kframe}
\begin{alltt}
\hlkwd{rmRecord}\hlstd{(d)}
\hlstd{plt} \hlkwb{<-} \hlkwd{findRecords}\hlstd{(}\hlstr{"smooth"}\hlstd{,} \hlkwc{ret_type} \hlstd{=} \hlstr{"id"}\hlstd{)}
\hlstd{plt}
\end{alltt}
\begin{verbatim}
## character(0)
\end{verbatim}
\end{kframe}
\end{knitrout}

\subsection{Processing and tracking Rmd-based dynamic reports}

Beyond individual artifacts, \rpkg{trackr} supports annotating and
tracking entire dynamic reports. We provide the
\rfun{knit\_and\_record} function, which wraps functionality in
\citet{xie_dynamic_2015}'s \rpkg{knitr} for generating dynamic reports. 

The \rfun{knit\_and\_record} function performs five high-level actions, it:
\begin{enumerate}
  \item{invokes \rpkg{knitr} to weave the input document into a dynamic report,}
  \item{captures and annotates all artifacts displayed within the
      report to a temporary backend,}
  \item{generates a metadata record for the report itself,}
  \item{associates the individual results and report by id, and finally}
  \item{records the report and individual results to its current backend}
\end{enumerate}

Once this is done, users can search for either the report as a whole
or a particular artifact within it. In either case, the metadata
associated with the record they find allows them to easily retrieve
the other.

\subsection{Tracking history to enhance annotations}

We have created the \rpkg{histry} \proglang{R} package for the automatic capture
and retention of the history of successfully-evaluated expressions. The 
\rpkg{trackr} package leverages \rpkg{histry} automatically to ensure code is 
available when recording objects, guaranteeing we can extract the types of 
code-based metadata discussed in \ref{aspects_of_code}.

\rpkg{histry} defines a formal mechanism for tracking expression-history 
generally, and provides two history trackers which are activated automatically
when the package is loaded. The first retains all \emph{successfully} evaluated 
top-level expressions, while the second tracks code evaluated while weaving 
dynamic reports using \rpkg{knitr} or \rpkg{rmarkdown}. 
\citep{xie_dynamic_2015,rmarkdown} Other trackers could be developed for use in 
other contexts, e.g., for use with \citet{Leisch_2002}'s \rpkg{Sweave} package.

\subsection{Customizing metadata extraction}

Different pieces of metadata can be extracted from different classes
of \proglang{R} object, as we pointed out in \ref{art_metadata}. We
provide specialized metadata-extraction capabilities for a number of
common artifact types, including plots, \rexpr{data.frames}, and
linear model fits.
% ML: maybe we should say "For other types of objects, we extract..."
% and actually give a condensed list of things we extract.
% And then remove the sentence below.
We also include a basic
metadata-extraction routine that acts as the default.

Beyond this, users and package developers customize metadata for a
particular class of \proglang{R} object in two ways: customizing the
tags extracted from objects of that class, and defining new formal
fields associated with metadata for that class and
% ML: This 'custom method' seems redundant with the first way
% but probably I am just confused. Should clarify this if feasible
% or just say that users can extend the metadata extraction for
% specific types of objects and just refer to the vignette.
creating a custom
method which extracts values for them. We refer interested readers to
the \rpkg{trackr} package's \rexpr{Extending-trackr} vignette for
details on how this is done.

\subsection{Storage backends}

Different ways of storing the metadata extracted by \rpkg{trackr} are best 
suited to different use-cases. Out-of-the-box, we support a light-weight JSON 
file-based backend and a heavier-weight Solr backend based on \citet*{rsolr}'s 
\rpkg{rsolr} package. The former is useful for tracking the analysis artifacts 
of a single analyst on a single machine, while the latter can support tracking
analyses of entire research-groups or departments. We refer readers to the 
relevant \rpkg{trackr} package vignette for instructions on setting up solr for
use with \rpkg{trackr}. 

Beyond these two backends, \rpkg{trackr} is fully extensible, allowing users to
create new backends that fit their computing needs. \rpkg{trackr} interacts
with backends exclusively via S4 generics that dispatch on the class of the
backend in use. The user can implement a backend by defining a class and 
providing methods for those generics. For more details and a working example of
a custom backend, see the \rpkg{trackr} vignette \rexpr{Extending-trackr}.

\subsection{Graphical frontends}

Many types of artifacts, particularly plots, have intuitive visual 
representations, so it is often desirable to search the artifact database 
through a graphical query interface. We have prototyped several graphical 
frontends for discovering and retrieving results recorded via \rpkg{trackr},
with different organizational scales and use cases in mind. When tracking an 
individual analyst's artifacts, the key features are automatic organization and
easy retrieval.  When multiple analysts are using \rpkg{trackr} within an 
organization and want to share and discover results across the group, there is 
a greater emphasis on search, discovery, and dissemination of artifacts, and we
need to adapt the interface accordingly. Finally, real-time collaboration would
benefit from a graphical live stream of results.

\subsubsection{Analysis-embedded Frontend}

\begin{figure}[h!]
\begin{center}
\includegraphics[width=0.7\columnwidth]{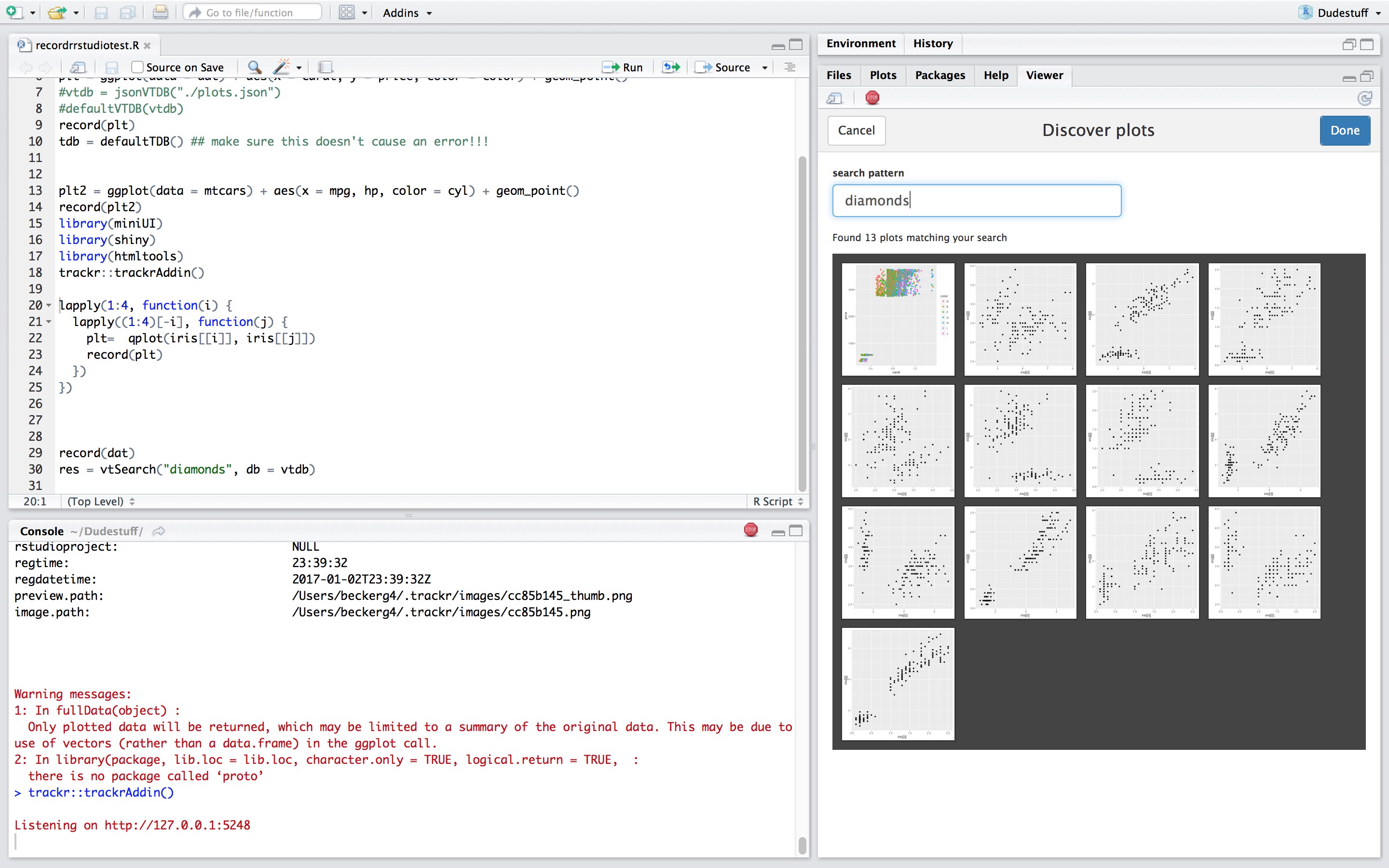}
\caption{{With the RStudio Addin version of our Shiny-based frontend, analysts
    can search for and view results seamlessly without leaving the RStudio IDE.%
}}
\label{fig:rstudio_addin}
\end{center}
\end{figure}

To incorporate \rpkg{trackr} into the analysis process, we provide a Shiny-based
\citep{shiny} frontend for discovery tasks within the analysis workflow. The
Shiny application can serve as an addin for the RStudio IDE (Figure \ref{fig:rstudio_addin}), directly 
integrating result discovery with result generation. The app accepts a simple 
text-based query and displays a gallery of matching results.

The result gallery may be a useful visualization in its own
right. When viewed as a single meta-visualization, the body of results
generated while analyzing a particular dataset can enable inference
about higher-order relationships that are not immediately obvious from
any individual result.

\subsubsection{Collaborative Frontend}

Blacklight \citep{blacklight} is an image-gallery frontend for Solr designed to
provide a visual catalog of assets for libraries and museums. Many of the 
features it offers out of the box align well with our goals of metadata-based
search, discovery, and visual exploration. These include automatic GUI controls
for faceting and restriction based on particular metadata fields, flexible
paginated display of result thumbnails, bookmarking and query history tracking,
and Solr's powerful search syntax and capabilities. 

\begin{figure}[h!]
\begin{center}
\includegraphics[width=0.7\columnwidth]{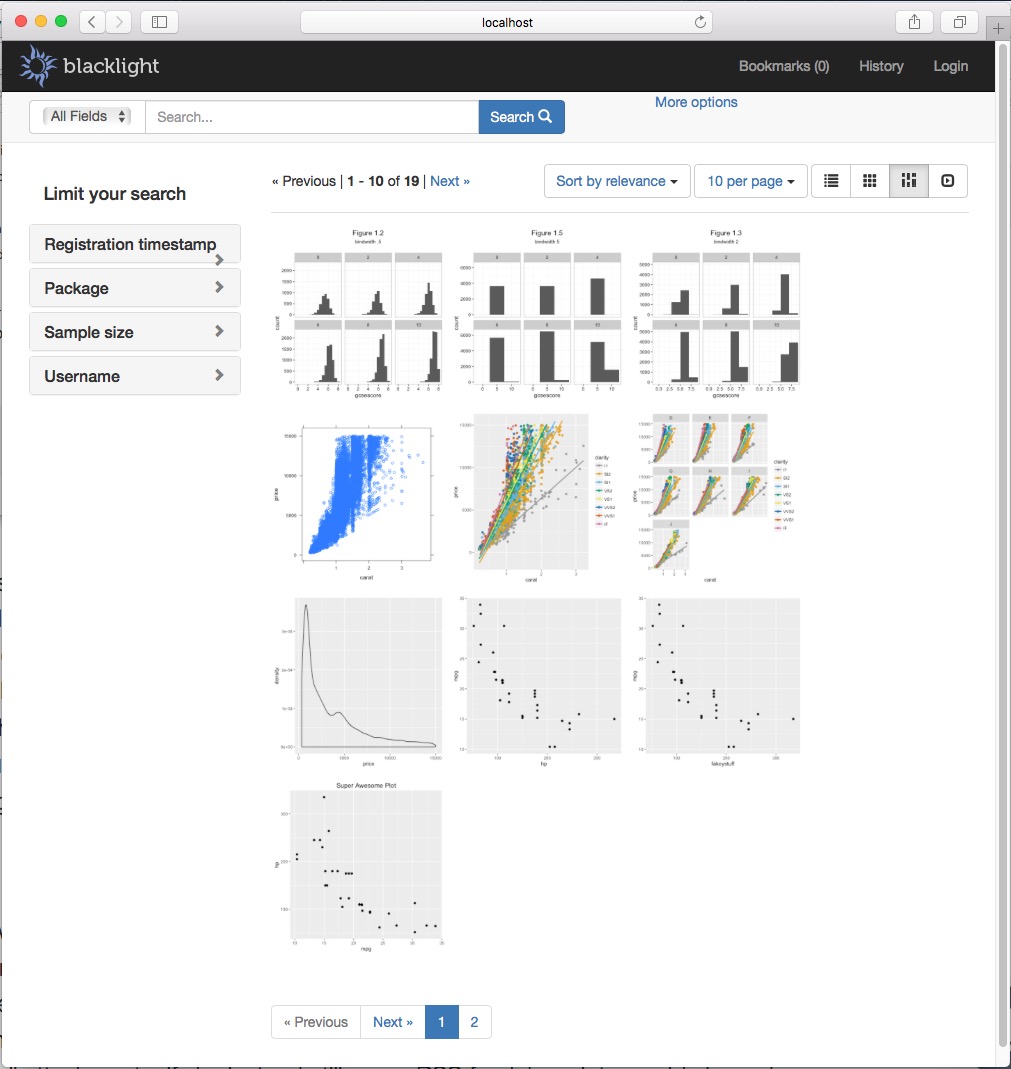}
\caption{{This blacklight gallery is running on top of a Solr core --- populated
    by \rpkg{trackr} ---     which contains visualizations of the
    \rexpr{diamonds}, \rexpr{Chem97}, and \rexpr{mtcars} datasets.%
}}
\label{fig:blacklight1}
\end{center}
\end{figure}

\begin{figure}[h!]
\begin{center}
\includegraphics[width=0.7\columnwidth]{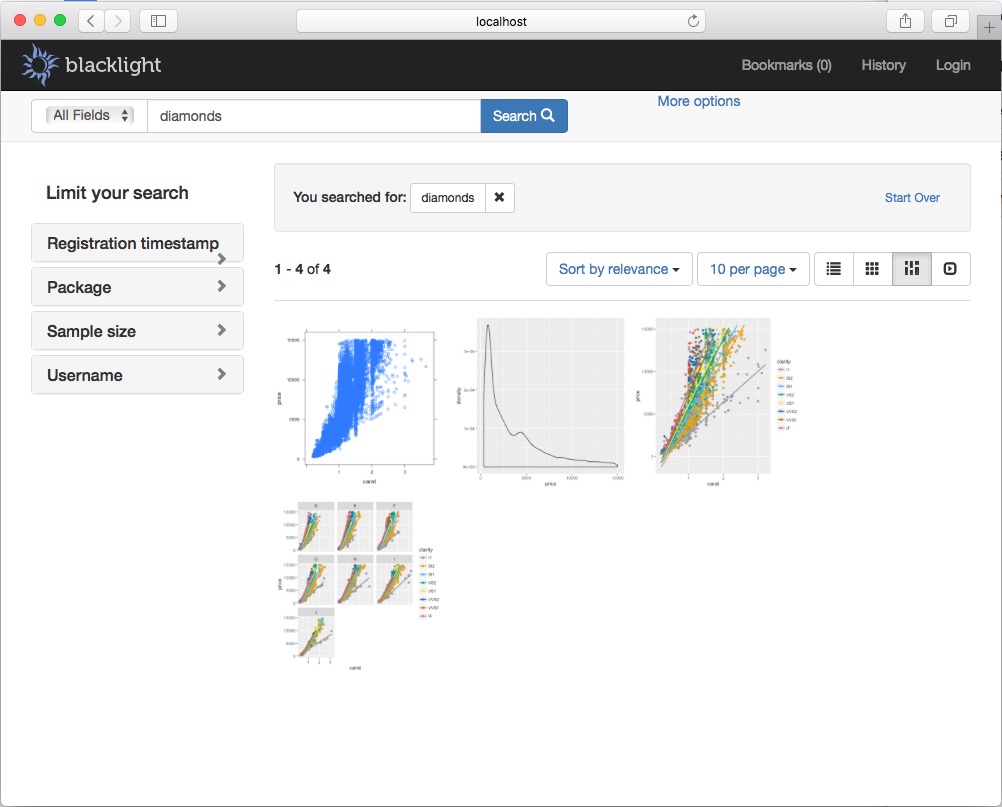}
\caption{{We use the Blacklight GUI to identify plots from our corpus which 
    display the \rexpr{diamonds} data. We could also do this by querying the 
    underlying Solr core directly, or via \rpkg{trackr}'s \proglang{R} API.%
}}
\label{fig:blacklight2}
\end{center}
\end{figure}

In Figures \ref{fig:blacklight1} and \ref{fig:blacklight2}, we see an example blacklight a gallery, and the result
of searching within it for plots created using the \rexpr{diamonds} dataset.

\subsubsection{Result Feeds}

When collaborating in real-time, scientists need to continually communicate
their results to the rest of the group.  The web technology Really Simple 
Syndication (RSS) is a standard mechanism for disseminating streams of content.
With a standard web browser, users can view an RSS feed of updates to the result
database, subject to a filter (conveniently encoded in the URL) that might 
restrict artifacts to plots made by a certain user, or marked with a certain 
tag. The RSS feed is essentially an auto-updating search, referenced by a URL. 
Even in the single user context, the RSS feed is useful for observing the 
history of an interactive analysis session. It is relatively straightforward to
configure Solr to provide an RSS feed of updates.

\section{Discussion}

\subsection{Prior Work}
Examples of attaching metadata to computational artifacts are found in
three related contexts: history-capture and provenance tracking within
scientific visualization platforms \citep{grasparc,Heer_2008,
  Woodruff_1997}, provenance capture in scientific-workflow systems
\citep{taverna,vistrails}, and data sharing and archival efforts
\citep{Michener_1997,dataone}.  Here the metadata primarly describe
how the artifact was created and, in the case of datasets, what data
they contain. These metadata are typically intended to enhance or
ensure reproducibilty of a plot or visualization state. In the context
of interaction history, this granularity can range from formally
modeled high-level decisions, as in GRASPARC, to a series of
individual low-level user interactions such as mouse clicks and menu
selections, with many systems falling somewhere in the middle
\citep{Heer_2008}. In the case of general computational provenance,
most specifications are relationship-based, tracking low-level details
about the data, actions and entities involved in generating a dataset
or result \citep{Moreau_2011,Missier_2013}.

\citet{Lee_1995} propose extending metadata capture to include the
querying and manipulation of the underlying data. They further propose
high-level summaries of both queries and query results, as well as
their use in modeling and inferring information about an analyst's
data explorations. While their ExBase system provided this
functionality only in the context of highly constrained,
database-backed visualization systems, their ideas provide a core
pillar of our work. Searchability of stored metadata has also come to
be considered a core feature of modern provenance systems
\citep{freire}, generalizing the ideas of \citet{Lee_1995}'s vision
and providing further foundation for the work we present here.

Finally, multiple services exist for sharing and publishing data
visualizations and computational artifacts tagged with
metadata. Figshare \citep{figshare} is the seminal service in this
space, supporting both visualizations and general artifacts, while
Plotly \citep{plotly} is a more recent addition focused on interactive
data visualizations. Despite recognizing the importance of good
metadata \citep{figshare_faq}, however, neither service annotates
artifacts automatically beyond basic logistics of the
submission. These services could provide alternative endpoints to our
framework, allowing users to supply some or all of the metadata we
generate for an artifact during submission. Similarly, metadata
extracted by \rpkg{trackr} could be used to construct submissions to
provenance stores such as ProvStore \citep{provstore}.

\subsection{Limitations}
\subsubsection{Relying on internal representations of plot objects}
Plot objects generated via the \rpkg{ggplot2} and \rpkg{lattice} frameworks are 
intended primarily for internal consumption by those frameworks. As such, those 
packages do not provide official, supported APIs to extract the types of 
information laid out in Section \ref{meth_plot_des_constr}, so we rely on 
implementation details that are subject to change. This incurs a substantial 
maintenance cost. In fact, \citet{ggplot} released version 2.0.0 of the 
\rpkg{ggplot2} package during the preparation of this manuscript, rendering 
this limitation non-hypothetical.

\subsubsection{Manual recording of artifacts}

\rpkg{trackr} requires an analyst to manually record artifacts in most
cases. This can be tedious, and may overlook intermediate artifacts
an analyst does not foresee needing in the future. This represents an
unfortunate but ultimately necessary trade-off between effective
discoverability and fully capturing the research process. 

An early prototype of \rpkg{trackr} contained an option to
automatically record every plot drawn to a graphics device whenever it
was loaded. We quickly found this approach to be infeasible.
During data exploration, analysts often generate many
similar versions of a plot
or model fit before arriving at the ``final'' version of that
artifact. These small iterations on the same artifact will typically
have very similar, overlapping metadata, which can make it difficult
to write simple discoverability queries that identify and return only
the desired final version.

% ML: this distinction between 'recording' and 'saving' is something
% that could be spelled out at the beginning of methods, just to
% lay the groundwork for the discussion of metadata extraction.
We view recording an artifact as an extension of saving it. Saving an
object is typically a manual step that implies importance, and the
same applies to recording. By explicitly recording an object, the
analyst declares the result to noteworthy enough to warrant later
retrieval.

The exception is dynamic report generation, where we automatically
capture every artifact displayed in the output, based on the principle
that if an artifact is important enough to be displayed in the
report, it is important enough to record.

\subsection{Future Directions}
\subsubsection{Automatic classification of plot type} \label{pred_plot_type}

Users may often want to restrict their searches to a particular class of plot 
--- histograms, residual scatterplots used for model diagnostics, gene 
expression heatmaps, etc. They may be searching for a particular plot, and know 
it falls in that class, or may wish to discover and explore all plots of that 
class meeting other search criteria. For some plot types, this is relatively 
straightforward using the metadata we already generate. Generally speaking, 
however, requiring users to hand-construct search terms which indicate plot 
classes ranges from burdensome to infeasible, depending on the type of plot in 
question.

One major avenue of possible future work is automatically annotating plots with 
predicted plot classes and other high-level semantic inferences about their 
content or purpose. The low-level, plot-specific metadata we generate now could 
be used as inputs to a combination of hand-constructed heuristics, and 
statistical learning and text mining methods, in order to generate such 
predictions.

\subsubsection{Extension to interactive graphics}

Our current framework supports only static graphics, and it would be natural to 
extend support to interactive graphics. The specification of an interactive 
graphic, independent of a specific state, encodes a great deal of
information in the modes of interactivity it offers. Tooltips are obviously 
useful as a direct visual representation of annotations, while 
interactive controls based on a variable or entered value encode an assumption 
that altering those values will change the content plotted and/or the 
conclusions the viewer might draw from it.

A snapshot of an interactive plot, particularly a state indicated by the user 
as final or particularly useful, also contains meaningful information. The 
parameters chosen, the subset of the data in view, the current selection, etc,
are all sources of inference about the question asked and answer found by the 
analyst. Moreover, freezing and preserving the state of an interactive graphic, 
to the extent possible, would contribute to the reproducibility and 
discoverability of those answers.

\subsubsection{Transforming static graphics into interactive ones}

In addition to \emph{annotating} interactive graphics, the metadata we capture 
about visualizations could also be used to \emph{create} them. Given two plots 
, we could, for example, use their metadata to infer that they plot the 
same dataset, and then create a linked-plot combining the two. This comes as a 
natural extension to the holistic view of related plots use-case \rpkg{trackr} 
already facilitates. 

Individual static graphics can also be (re-)rendered as interactive graphics 
that provide GUI controls for setting particular parameters used by the code 
\citep{idyndocs,becker2014dynamic}. A plot that displays the subset of data 
from a particular year, for example, might be re-rendered with a year-selector 
via the combination of a straightforward transformation of the code and an 
engine to redraw the plot. With \rpkg{trackr} ensuring code is available, 
scientists could interactively explore shared results, even when the original 
analyst created only static data visualizations.

\subsubsection{Annotating the analysis itself}

It is rare that a single, independent result adequately represents the output 
of an entire analysis. In our experience, analyses typically produce many 
artifacts, each of them connected to the others through the thread of the 
analysis process. To capture these relationships, we need to track the analysis 
as a whole. 

An annotated analysis might simply be a collection of annotated computational 
artifacts, along with the full code and data implementing the analysis. This 
collection would be annotated with analysis-level metadata. Some of these 
top-level annotations would likely be inherited directly from among the 
annotations of the artifacts, particularly if the system is able to identify 
(or is told) which artifacts are representative. Other annotations could be 
inferred from the number and type of artifacts the analysis generated, from 
combinations of annotations across artifacts, and from the full body of code 
that performed (and would re-perform) the analysis.

\section{Conclusion}

We have presented the \rpkg{trackr} framework, consisting of our
\rpkg{trackr} and \rpkg{histry} packages. Our framework provides three
capabilities which combine to improve reproducibility, understandability, and
discoverability of \proglang{R}-based results. First, it provides a
mechanism for standardized capture and storage of results annotated
with metadata. Second, it provides a mechanism for automatically
extracting meaningful metadata from the results themselves as well as
the code, environment, and data that generated them. Finally it
provides the ability to query and search for results, based on their
metadata, in multiple contexts via the presented proof-of-concept
interfaces.

By automatically organizing, annotating, and indexing results,
\rpkg{trackr} enables analysts and organizations to more easily find,
understand, reproduce and extend previous results, whether their own
or of others, accelerating the iterative research process.

\section{Availability}
Up-to-date versions of our \rpkg{trackr} and \rpkg{histry} packages
are available on Github at \url{http://github.com/gmbecker/recordr}
and \url{http://github.com/gmbecker/histry}, respectively. The
\rpkg{rdocdb} dependnecy is also available on Github at
\url{http://github.com/lawremi/rdocdb}; all other dependencies are
available on CRAN. After responding to anything that might arise
during review, they will be posted on CRAN.

\bibliography{converted_to_latex.bib}
  
\end{document}